\documentclass[10pt,twocolumn,letterpaper]{article}

\usepackage[pagenumbers]{cvpr} % To force page numbers, e.g. for an arXiv version

\usepackage{graphicx}
\usepackage{amsmath}
\usepackage{amssymb}
\usepackage{booktabs}

\usepackage[pagebackref,breaklinks,colorlinks]{hyperref}

\usepackage[capitalize]{cleveref}
\crefname{section}{Sec.}{Secs.}
\Crefname{section}{Section}{Sections}
\Crefname{table}{Table}{Tables}
\crefname{table}{Tab.}{Tabs.}

\def\cvprPaperID{4725} % *** Enter the CVPR Paper ID here
\def\confName{CVPR}
\def\confYear{2022}

\begin{document}

\title{Learning A 3D-CNN and Transformer Prior for Hyperspectral Image Super-Resolution}

\author{
Qing Ma$^1$ \hspace{0.2in} Junjun Jiang$^{1,2}$\thanks{Corresponding author: jiangjunjun@hit.edu.cn} \hspace{0.2in} Xianming Liu$^{1,2}$  \hspace{0.2in} Jiayi Ma$^{3}$\\
%$^1$School of Computer Science and Technology, Harbin Institute of Technology, Harbin 150001, China\\
%$^2$Electronic Information School, Wuhan University, Wuhan 430072, China
$^1$Harbin Institute of Technology \hspace{0.2in}
$^2$Peng Cheng Laboratory \hspace{0.2in}
$^3$Wuhan University
}

\maketitle

%%%%%%%%% ABSTRACT
\begin{abstract}
  To solve the ill-posed problem of hyperspectral image super-resolution (HSISR), an usually method is to use the prior information of the hyperspectral images (HSIs) as a regularization term to constrain the objective function. Model-based methods using hand-crafted priors cannot fully characterize the properties of HSIs. Learning-based methods usually use a convolutional neural network (CNN) to learn the implicit priors of HSIs. However, the learning ability of CNN is limited, it only considers the spatial characteristics of the HSIs and ignores the spectral characteristics, and convolution is not effective for long-range dependency modeling. There is still a lot of room for improvement. In this paper, we propose a novel HSISR method that uses Transformer instead of CNN to learn the prior of HSIs. Specifically, we first use the proximal gradient algorithm to solve the HSISR model, and then use an unfolding network to simulate the iterative solution processes. The self-attention layer of Transformer makes it have the ability of spatial global interaction. In addition, we add 3D-CNN behind the Transformer layers to better explore the spatio-spectral correlation of HSIs. Both quantitative and visual results on two widely used HSI datasets and the real-world dataset demonstrate that the proposed method achieves a considerable gain compared to all the mainstream algorithms including the most competitive conventional methods and the recently proposed deep learning-based methods. %The code is available at https://anonymous.4open.science/r/cvpr2022-4725.
\end{abstract}

%%%%%%%%% BODY TEXT
\section{Introduction}
\label{sec:intro}

Hyperspectral images (HSIs) have high spectral resolution and are widely used in various computer vision tasks, including target recognition and tracking \cite{pan2003face,van2010tracking}, medical image processing \cite{akbari2012hyperspectral}, and remote sensing \cite{fauvel2012advances,camps2013advances}.
While the HSI can achieve a high spectral resolution with contiguous and narrow bands, its spatial resolution is usually much coarser than that of the RGB images in our daily life. This is due to the fact that the dense spectral bands in the hyperspectral sensors make a limited amount of photons reached one narrow spectral window averagely. The low spatial resolution of the HSI captured by the sensor greatly limits its value for application. A natural solution is to instead capture a low-resolution (LR) HSI and a high-resolution (HR) multispectral image (MSI), \emph{e.g.}, RGB image, and fuse them into a resultant image with high spatial and spectral resolution simultaneously. This procedure is referred to as hyperspectral multispectral image fusion (HS/MS fusion) or hyperspectral super-resolution.

Traditional HS/MS fusion methods employ prior knowledge of HSIs as regularizer to solve such a seriously ill-posed problem, \emph{e.g.}, sparse representation \cite{akhtar2015bayesian, kawakami2011high, dong2016hyperspectral}, low-rank prior \cite{veganzones2015hyperspectral, zhang2016multispectral},  non-local \cite{dian2017hyperspectral, zhang2018exploiting, xu2019nonlocal} and self-similarity \cite{han2018self}. They have achieved better and better performance, because these priors are getting closer and closer to the essential characteristics of the data. However, the prior knowledge is predefined manually and is born with the following shortcomings. On the one hand, a high level of wisdom is required. On the other hand, these hand-crafted priors have limitations and often fail to reflect all the characteristics of the data. These severely restrict the performance and generalization ability of the optimization-based methods mentioned above.

Given the recent advances in deep learning, many convolutional neural networks (CNNs) based image super-resolution approaches have garnered attention recently. To solve the above problems of traditional HS/MS fusion methods, some researchers use deep learning to cope with the fusion based HSISR problem. Unlike traditional methods of manually designing a prior, these deep learning-based methods embrace a large amount of data to learn data-driven priors, greatly improving the fusion performance. Researchers have proposed some specialized modules, such as spatial attentions, channel attentions, and joint ones, to exploit the spatial, spectral, or spatio-spectral priors of HSIs.
To take advantage of the observation model and address the gap between the optimization-based and learning-based methods, most recently, a series of HS/MS fusion methods based on unfolding networks have been proposed. For example, Xie \emph{et al.} \cite{xie2020mhf} proposed an effective unfolding network based on MHF-net, which first constructs an MS/HS fusion model and then builds the proposed network by unfolding the proximal gradient algorithm to solve the proposed model. Wang \emph{et al.} \cite{wang2019deep} proposed the DBIN model, where the estimation of the observation model and the fusion process is optimized iteratively and alternatively during the super-resolution reconstruction. Dong \emph{et al.} \cite{dong2021model} proposed an approach based on MoG-DCN, an iterative HSISR algorithm based on a deep HSI denoiser to leverage both domain knowledge likelihood and deep image prior. Essentially, these methods all try to cast the HS/MS fusion optimization problem into joint learning of the observation model and a deep denoiser prior. The intrinsic difference among the above three tasks lies in how to model the denoiser prior. MHF-net and DBIN stack several ResBlocks \cite{he2016deep} to model the data prior, while MoG-DCN leverages a relatively complex U-net \cite{ronneberger2015u} to exploit the prior.

In this paper, we try to design a more powerful network that can fully extract the priors of HSIs. We observe that the HSIs are three-dimensional data cubes, and their priors can be divided into two aspects: spatial priors and spectral priors. We need to design appropriate networks for these two aspects to better learn the priors of HSIs. We propose to use Transformer \cite{vaswani2017attention} to learn the spatial prior of HSIs. The self-attention mechanism in the Transformer requires each pixel of the image to pay attention to each other pixel, so that Transformer has long-range modeling capabilities and has achieved good performance in many visual tasks. It is very suitable for extracting the spatial prior of HSIs. As for the spectral prior, 3D-CNN is a natural and effective choice. It can pay attention to the correlation of spatio-spectra while learning spectral priors. Keeping these in mind, we propose an HS/MS fusion model, namely 3D-CNN and Transformer prior network (3DT-Net). We first use the proximal gradient algorithm to solve the optimization problem with the observation models of LR-HSIs and HR-MSIs, and then we unfold the iterative process of the proximal gradient algorithm into a multistage network. Particularly, in each iteration, unlike previous learning-based methods that construct 2D-CNN to learn priors, we design a 3D-CNN and Transformer network with
Swin Transformer layers \cite{liu2021swin} to exploit spatial priors and 3D convolutional layers to exploit spatio-spectral correlation priors. Experimental results have shown that the proposed 3DT-Net outperforms many recently proposed HS/MS fusion methods.

%-------------------------------------------------------------------------

\begin{figure*}[t]
  \centering
  \includegraphics[width=0.95\textwidth]{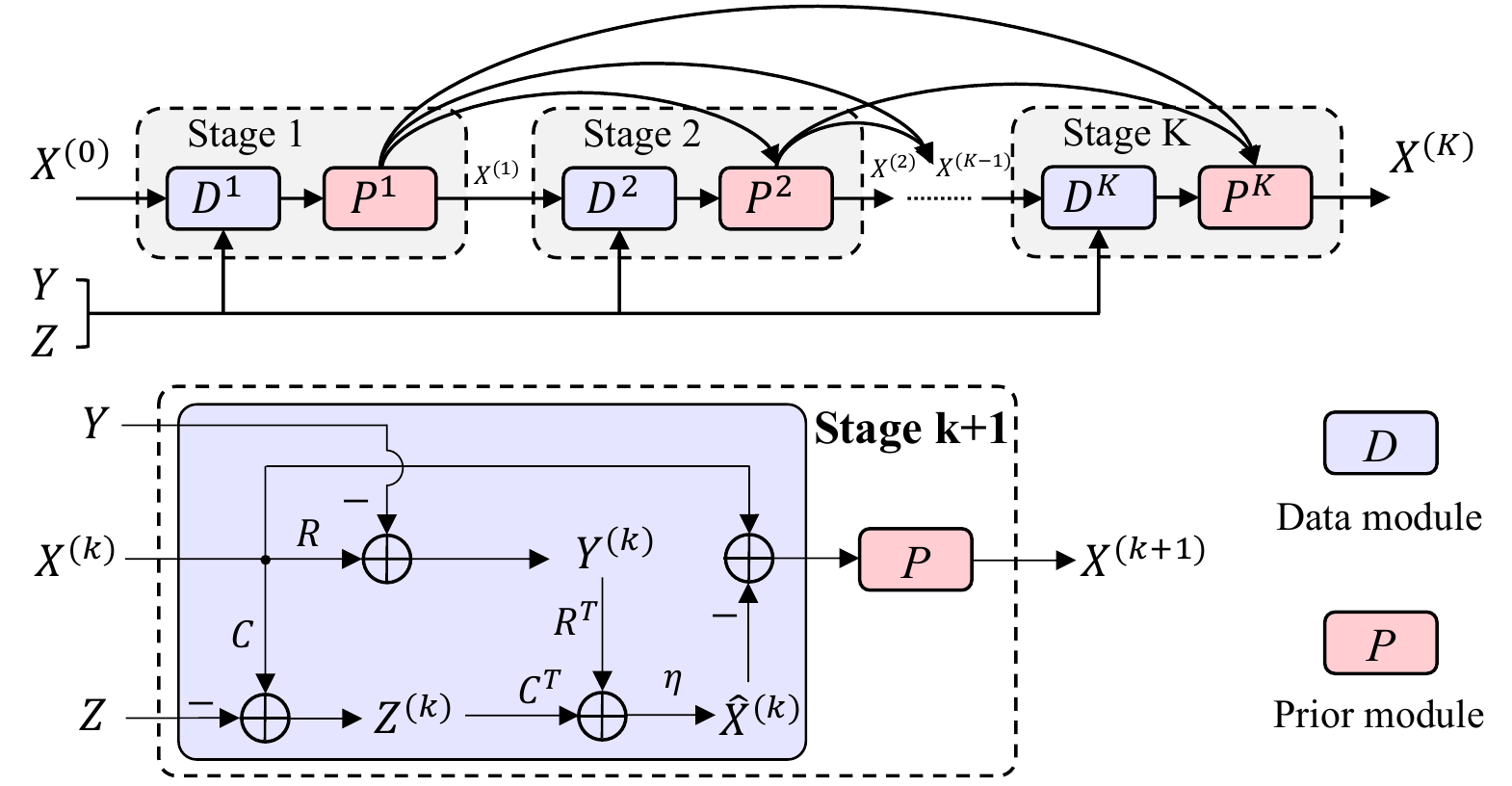}\\
  \caption{The flowchart of our proposed 3DT-Net deep unrolling framework.}\label{fig:framework}
  %\vspace{-0.3cm}
\end{figure*}

\section{Related Work}
\label{sec:related}

In this section, we introduce related existing methods, particularly focusing prior modeling in hyperspectral image super-resolution and vision Transformer.

\subsection{Prior Modeling in HSISR}
HSIs super-resolution is an ill-posed problem, a common solution is to choose a prior or regularization to constrain the optimization equation. Various priors have been already exploited to regularize the HSIs Super-resolution problem. In \cite{palsson2013new}, Frosti \emph{et al.} used total variation to regularize an ill-posed problem dictated by a widely used explicit image formation model. Considering that HSIs are redundant in nature, sparse prior can be used to constrain the spatial and spectral correlation of HSIs. Wei \emph{et al.} \cite{wei2015hyperspectral} designed a sparse regularization term, relying on a decomposition of the scene on a set of dictionaries. In \cite{akhtar2015bayesian}, Akhtar \emph{et al.} proposed a generic Bayesian sparse coding strategy to be used with Bayesian dictionaries learned with the Beta process. In addition, spectral unmixing prior \cite{yokoya2011coupled, wei2016multiband} and self-similarity \cite{han2018self} are also often used to address HSISR. The main drawback of these methods is the quality of the recovered HSIs highly depend on whether the selected prior is appropriate. In addition, it requires manual tweaking of its balancing parameters between the data term and prior term.

In recent years, deep learning have pushed forward the frontier of many computer vision tasks, including image classification and object detection \cite{he2016deep,girshick2014rich}, image retrieval \cite{qayyum2017medical}, and so on. Inspired by these representative work, using deep learning methods to accomplish the HS/MS fusion has attracted wide attention. Different from the traditional method of manual design a prior, the deep learning-based methods use a large amount of data to learn data-driven priors. Dian \emph{et al.} \cite{dian2018deep} combined traditional methods with deep learning methods. They first initialize the HR-HSI by solving a Sylvester equation, and then use deep CNN to learn the priors of the target HR-HSI. Xie \emph{et al.} \cite{xie2020mhf} and Wu Wang \emph{et al.} \cite{wang2019deep} used ResNet to learn HSIs priors. Dian \emph{et al.} \cite{dian2020regularizing} proposed using a CNN denoiser to regularize the HS/MS fusion model. In \cite{dong2021model}, Dong \emph{et al.} proposed using U-net to replace ResNet to learn a denoising prior for HS/MS fusion.

\subsection{Vision Transformer}
Transformer was proposed by Vaswani \emph{et al.} \cite{vaswani2017attention} for NLP. Since directly replacing the CNN convolution layers with Transformer layers to process images will bring a huge computational burden, the early methods are to augment a standard CNN model with Transformer layers \cite{bello2019attention, srinivas2021bottleneck, wang2018non, yin2020disentangled}. Recently, Chen \emph{et al.} \cite{chen2021pre} proposed a backbone model IPT for image restoration based on Transformer. However, IPT has a huge amount of parameters and calculations, and it is very difficult to train such a model. In order to solve the problem of large amount of calculation when using the Transformer to process image tasks, Dosovitskiy \emph{et al.} \cite{dosovitskiy2020image} proposed to split the training image into a sequence of image patches. However, \cite{dosovitskiy2020image} cannot handle high-resolution images, due to its low-resolution feature maps and the quadratic increase in complexity with image size. Follow up \cite{dosovitskiy2020image}, Liu \emph{et al.} proposed Swin Transformer \cite{liu2021swin}, a hierarchical Transformer whose representation is computed with Shifted windows. Swin Transformer can not only process high-resolution images like CNN, but also has the advantage of Transformer to model long-range dependency. In \cite{liang2021swinir}, an image restoration Transformer was developed based on Swin Transformer, which achieves state-of-the-art performance in various image reconstruction tasks.

%-------------------------------------------------------------------------
\section{Proposed Method}
\label{sec:proposed}
\subsection{Model Formulation}
In this paper, we use lowercase letters to denote scalars, use bold letters to denote matrices, and use curlycue to denote tensors. Specially, let $\mathcal{Y}\in\mathbb{R}^{W\times H\times s}$ represent the HR-MSI and $\mathcal{Z}\in\mathbb{R}^{w\times h\times S}$ represent the LR-HSI. The goal of HSISR is to combine information coming from an LR-HSI and an HR-MSI. The former has high spectral resolution, with $S$ spectral bands, but low geometric resolution with $w$ and $h$ being image width and height, respectively. The latter has high spatial resolution with $W$ and $H$ being image width and height, respectively, but a low spectral resolution $s$. We aim to estimate a fused image $\mathcal{X}\in\mathbb{R}^{W\times H\times S}$ with both high spatial and high spectral resolution.

If we reshape the spatio-spectral data-cube $\mathcal{Y}$, $\mathcal{Z}$ and $\mathcal{X}$ to their matrix formulations, \emph{i.e.}, $\boldsymbol Y \in\mathbb{R}^{WH\times s}$, $\boldsymbol Z \in\mathbb{R}^{wh\times S}$, and $\boldsymbol X \in\mathbb{R}^{WH\times S}$, the observation models for the HR-MSI and LR-HSI can be written as follows:
\begin{align}
    \boldsymbol Y &= \boldsymbol X \boldsymbol R,\\
    \boldsymbol Z &= \boldsymbol C \boldsymbol X.
\end{align}
where $\boldsymbol R \in \mathbb{R}^{S \times s}$ is the spectral response function of the multispectral sensor, $\boldsymbol C \in \mathbb{R}^{wh \times WH}$ represents the degradation operator which is often assumed to be composed of a cyclic convolution operator and a down-sampling matrix. We can obtain $\boldsymbol X$ by solving the following optimization problem
\begin{equation}
\begin{aligned}
\label{eq:target optimization}
      \mathop{\min}_{\boldsymbol X}\ \  \frac{1}{2}\lVert \boldsymbol X\boldsymbol R-\boldsymbol Y \rVert_F^2+ \frac{1}{2}\lVert \boldsymbol C \boldsymbol X -\boldsymbol Z \rVert_F^2+\lambda f(\boldsymbol X).
\end{aligned}
\end{equation}
where the first and second terms are fidelity terms and the third term is a regularization term which represents prior knowledge of $\boldsymbol X$. In the conventional HS/MS fusion methods, most of the priors are hand-crafted based on empirical observation, such as sparse priors \cite{kawakami2011high, dong2016hyperspectral}, low-rank priors \cite{veganzones2015hyperspectral, zhang2016multispectral}, self-similarity \cite{han2018self}. Although they have achieved promising results, but they only use a certain property of hyperspectral images, and it is difficult to apply to a variety of hyperspectral images. Pioneering works \cite{xie2020mhf, wang2019deep, dong2021model} have proved that the performance of data-driven priors exceeds that of hand-crafted priors. Therefore, in this work, we use a network to learn the data-driven implicit prior.

Since the specific form of the prior is not specified, the regularization term is usually non-smooth and non-differentiable functions of the outputs. Therefore, in this paper we use a proximal gradient algorithm \cite{beck2009fast} to solve Eq. (\ref{eq:target optimization}). Let $\mathcal L(\boldsymbol X) = g(\boldsymbol X) + \lambda f(\boldsymbol X)$, where $g(\boldsymbol X)$ is differentiable (includes the first two terms of Eq. (\ref{eq:target optimization})) and $\lambda f(\boldsymbol X)$ is non-differentiable. The proximal gradient algorithm minimizes Eq. (\ref{eq:target optimization}) by iterating the following equation until convergence:
\begin{equation}
\begin{aligned}
\label{eq:optimization function}
    \boldsymbol X^{(k+1)}&=\text{prox}_{\lambda \eta f}(\boldsymbol X^{(k)} - \eta \bigtriangledown g(\boldsymbol X^{(k)})),
\end{aligned}
\end{equation}
where $\eta$ plays the role of step size, and $\text{prox}_{\lambda \eta f}(\cdot)$ is a proximal operator dependent on $\lambda$, $\eta$ and $f$. Since $\bigtriangledown g(\boldsymbol X^{(k)}) = (\boldsymbol X^{(k)} \boldsymbol R - \boldsymbol Y) \boldsymbol R^T + \boldsymbol C^T (\boldsymbol C \boldsymbol X - \boldsymbol Z)$, we substitute it into Eq. (\ref{eq:optimization function}) to obtain the following optimization problem that can be solved iteratively:
\begin{footnotesize}
\begin{equation}
\begin{aligned}
\label{eq:final optimization function}
    \boldsymbol X^{(k+1)} &=\text{prox}_{\lambda \eta f}(\boldsymbol X^{(k)} - \eta (\boldsymbol X^{(k)} \boldsymbol R - \boldsymbol Y) \boldsymbol R^T + \boldsymbol C^T (\boldsymbol C \boldsymbol X^{(k)} - \boldsymbol Z)).
\end{aligned}
\end{equation}
\end{footnotesize}

%-------------------------------------------------------------------------
\subsection{Optimization-based Unfolding Network}
Conventional proximal gradient algorithm requires many iterations to converge and is computationally expensive. Moreover, the step size $\eta$ has to be selected by hand and it is difficult to find the optimal $\eta$. To overcome these shortcomings, recent works \cite{dong2018denoising, xie2020mhf, dong2021model} used a deep unfolding network to unfold the iterative optimization steps into a series of networks. These models have been trend-setting and promising in solving the inverse problems. The methods based on deep unfolding strategy have the fundamental advantages of interpretability, fewer parameters, and fast convergence. We derive our motivation from these works and design a deep unfolding network to solve the optimization Eq. (\ref{eq:final optimization function}).

Before solving Eq. (\ref{eq:final optimization function}), we first present the physical meaning of matrix operations in this paper: the left multiplication matrix represents the spatial transformation of the input, and the right multiplication matrix represents the channel transformation of the input. Then we unfold the iteration Eq. (\ref{eq:final optimization function}) into the following equivalent four sequential parts:
\begin{align}
    &\boldsymbol Y^{(k)} = \boldsymbol X^{(k)} \boldsymbol R - \boldsymbol Y, \label{eq:Yk} \\
    &\boldsymbol Z^{(k)} = \boldsymbol C \boldsymbol X^{(k)} - \boldsymbol Z, \label{eq:Zk} \\
    &\hat{\boldsymbol X}^{(k)} = \eta (\boldsymbol Y^{(k)} \boldsymbol R^T + \boldsymbol C^T \boldsymbol Z^{(k)}), \label{eq:X_hat} \\
    &\boldsymbol X^{(k+1)} = \text{prox}_{\lambda \eta f}(\boldsymbol X^{(k)} - \hat{\boldsymbol X}^{(k)}), \label{eq:Xk}
\end{align}
where in Eq. (\ref{eq:Yk}), $\boldsymbol R \in \mathbb{R}^{S \times s}$ represents channel decrease operator, which can be performed by using some convolution layers. The simplest way is to apply the  $1\times1$ convolution to decrease the channel number from $S$ to $s$. In this paper, we leverage the more flexible $3\times3$ convolution layer to achieve this. In Eq. (\ref{eq:Zk}), $\boldsymbol C \in \mathbb{R}^{wh \times WH}$ represents the spatial downsampling operator, we perform it by several $2\text{D}$ convolution layers with kernel stride 2 and thus each $2\text{D}$ convolution layer carrying out $2$ times downsampling. $\boldsymbol R^T \in \mathbb{R}^{s \times S}$ in Eq. (\ref{eq:X_hat}) represents channel increase operator, we perform it by using the $3\times3$ convolution layers in a similar vein. $\boldsymbol C^T\in \mathbb{R}^{HW \times wh}$ represents an upsampling operator, and we perform it by several $2\text{D}$ transposed convolution layers \cite{dumoulin2016guide} with each $2\text{D}$ transposed convolution layer carrying out $2$ times upsampling. Since it is almost impossible to find the best $\eta$ by hand, we set $\eta$ to be a learnable parameter and update it as the backward propagation of errors. In Eq. (\ref{eq:Xk}), $\text{prox}_{\lambda \eta f}(\cdot)$ is a proximal operator used to model the image prior. For the simple prior modeling, such as the $l$-norm regularization for sparse representation, the proximal is tractable and can be modeled in closed-form through soft-thresholding and shrinkage based unrolling network \cite{zhang2018ista}. More details about the prior network will be introduced in the next section.

Motivated by the proximal gradient iterations in Eq. \eqref{eq:Yk}-\eqref{eq:Xk}, we can now perform these invert linear inverse tasks by efficient network architectures. Fig. \ref{fig:framework} shows the architecture of the proposed network. The observations $Y$ and $Z$ forms the inputs for all iterations, which together with the output $X^{(k)}$ of the previous iteration form the variable for the current iteration. The network contains $K$ iteration stages, and each iteration stage is composed of a data module (corresponding to Eq. \eqref{eq:Yk}-\eqref{eq:X_hat}) and a prior module (corresponding to Eq. \eqref{eq:Yk}-\eqref{eq:Xk}), which is modeled via an efficient network architecture, as will be elaborated in the next section. In the first stage (\emph{i.e.}, $k=1$), we perform Bicubic upsampling on LR-HSI $Z$ to initialize $\boldsymbol X^{(0)}$. To reduce the number of parameters, we share the parameters for different stages. Therefore, the proximal map will be modeled by unrolling the proximal gradient iterations through a recurrent-like architecture \cite{mardani2018neural}. To avoid information loss during transmissions between iterations, inspired by some previous works \cite{dong2021model, wu2021dense} we propose to densely connect the hierarchical feature maps from the previous prior module to the subsequent prior modules, as shown in Fig. \ref{fig:framework}.

% make full use of the hierarchical features and alleviate the problem of vanishing gradient, we connect those feature maps from the previous prior module to the subsequent prior modules. The feature maps of the first $2D$ convolution layer in prior module are connected to the first $2D$ convolution layers of the subsequent prior modules, respectively.

\begin{figure*}[t]
  \centering
  \includegraphics[width=0.95\textwidth]{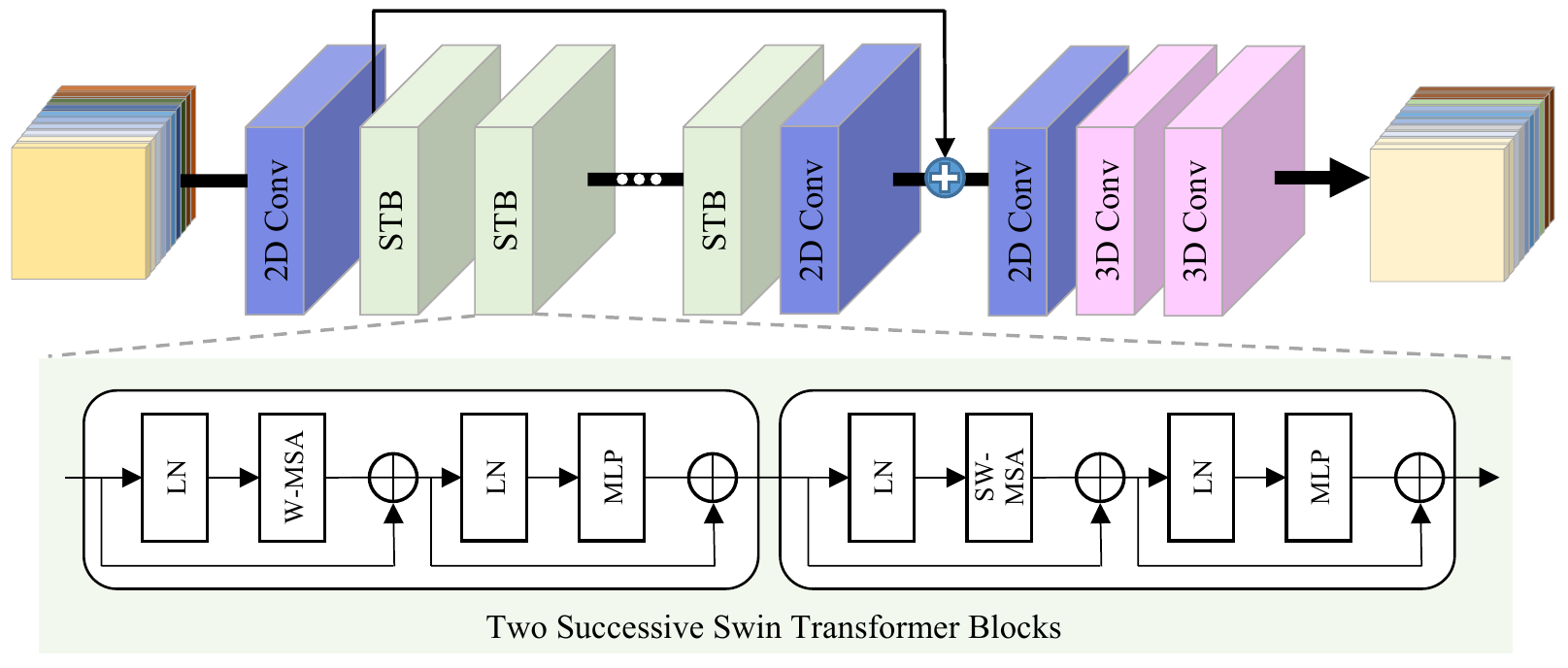}\\
  \caption{Illustration of the network architecture of the proposed 3D-CNN and Transformer Prior (3DT) module. LN and MLP, are the LayerNorm and MultiLayer Perceptron layers, respectively. W-MSA and SW-MSA represent the window based multi-head
self-attention using regular and shifted window partitioning configurations, respectively.}\label{fig:prior}
  %\vspace{-0.3cm}
\end{figure*}

%-------------------------------------------------------------------------
\subsection{3D-CNN and Transformer Prior Network}
For these deep unfolding networks, the proximal operator is usually modeled as a deep neural model, which resembles the projection from the (noisy) observation onto the manifold (where the true image lies in) of visually plausible images. Therefore, the proximal operator can be seen as a denoiser that gradually removes the artifacts from the noisy observation.
%, and designing an effective denoiser becomes especially important.
Pioneering works (\emph{e.g.}, DBIN \cite{wang2019deep}, MHF-net \cite{xie2020mhf}, MoG-DCN \cite{dong2021model} \emph{etc}.) usually used a sub-network to learn this operation. In DBIN \cite{wang2019deep} and MHF-net \cite{xie2020mhf}, the authors adopted ResNet to learn this proximal operator and attained satisfactory performance. Recently, MoG-DCN \cite{dong2021model} proposed to use U-net \cite{ronneberger2015u} as the backbone network architecture to learn this proximal operator which achieves the SOTA performance on the HS/MS fusion task. From these works, it is easy to analyze that the performance of the selected prior estimation network is the essential factor that determines the performance of the entire HS/MS fusion network. Therefore, in this paper, we  take the existing method one step further and propose a new denoiser prior learning network based on 3D-CNN and Transformer network, which introduces the powerful Transformer \cite{vaswani2017attention} structure into the prior learning network.

The architecture of the 3D-CNN and Transformer prior network is illustrated in Fig. \ref{fig:prior}. The main components of the 3D-CNN and Transformer prior network are Swin Transformer layer (STL) and 3D convolution layer. The STL was proposed for the first time in \cite{liu2021swin}. It has long-range modeling capabilities like the original Transformer, and the amount of calculations is much smaller than that of the original Transformer. Therefore, it is a very good choice for learning data priors. However, STL is designed for natural RGB images. It fully considers the spatial self-similarity of the image, but does not consider the spectral correlation of the HSI data. Therefore, it is not appropriate to directly use STL to learn the priors of 3D HSIs. To address this problem, we use 3D convolutional layers to learn the spectral priors of HSIs. Compared with filters in 2D convolution, filters in 3D convolution are more flexible in the channel dimension, making it more suitable for exploring the spatio-spectral correlation of HSIs.

%------------------------------------------------------------------------
\subsection{Training Details}
We resort to the $L_1$ loss to train our network
\begin{equation}
\begin{aligned}
    L &= \lVert \boldsymbol X^{(K)}-\boldsymbol X \rVert_{1},
\end{aligned}
\end{equation}
where $\boldsymbol X^{(K)}$ is the final output HR-HSI. $\boldsymbol X$ is the ground truth HR-HSI.

In our implementation, we initialize $\eta$ with a uniform distribution in $[0,1]$ and in each stage are updated with gradient-based method. We use Adam optimizer \cite{kingma2014adam} with $\beta_1=0.9$, $\beta_2=0.999$, $\epsilon=10^{-8}$ to train the network for 250000 iterations with a batch size of 8. We set each epoch to have 2500 iterations, thus there are 100 epochs. The learning rate is initialized as 0.0001. We implement and train our network using Pytorch framework with an NVIDIA Tesla V100 GPU.

%------------------------------------------------------------------------
\section{Experiments and Analyses}
\label{sec:Experiments}
In this section, we compare our proposed method with eight other state-of-the-art methods including four traditional methods, \emph{e.g.},
Hysure \cite{simoes2014convex}$\footnote{https://github.com/alfaiate/HySure}$,
NSSR \cite{dong2016hyperspectral}$\footnote{http://see.xidian.edu.cn/faculty/wsdong/HSI\_SR\_Project.htm}$,
CSTF \cite{li2018fusing}$\footnote{https://github.com/renweidian/CSTF}$,
LTTR \cite{dian2019learning}$\footnote{https://github.com/renweidian/LTTR}$,
 and four recently proposed deep learning-based methods, \emph{e.g.},
CNN-FUS \cite{dian2020regularizing}$\footnote{https://github.com/renweidian/CNN-FUS}$,
MHF-net \cite{xie2020mhf}$\footnote{https://github.com/XieQi2015/MHF-net}$,
DBIN \cite{wang2019deep}$\footnote{https://github.com/wwhappylife/Deep-Blind-Hyperspectral-Image-Fusion}$,
MoG-DCN \cite{dong2021model}$\footnote{https://github.com/chengerr/Model-Guided-Deep-Hyperspectral-Image-Super-resolution}$.
As for objective comparisons, we use four picture quality indices (PQIs) to evaluate the performance of different methods, including peak signal-to-noise ratio (PSNR), SAM \cite{yuhas1992discrimination}, erreur relative globale adimensionnelle de synth\`{e}se (ERGAS) \cite{wald2002data} and structure similarity (SSIM) \cite{wang2004image}. PSNR and SSIM are calculated on each 2D spatial image, which evaluate the similarity between the recovered HSI and the ground truth based on MSE and structural consistency, respectively. SAM calculates the average angle between spectrum vectors of the recovered HSI and the ground truth. ERGAS reflects the overall quality of the recovered HSI. Smaller values of ERGAS and SAM suggest better results, while larger values of PSNR and SSIM imply better results.

\begin{table}
%\vspace{1mm}
\begin{center}
%\vspace{1.5mm}
\normalsize
\setlength{\tabcolsep}{6.5pt}
\caption{Influence of the number of iterations for our 3DT-Net.}\label{tab:iterations}
\begin{tabular}{c|ccccc}
  \hline
  \hline
Model	&	PSNR	&	SAM	&	ERGAS	&	SSIM\\
\hline
3DT-Net ($K=1$)	&	49.35	&	2.40	&	0.53	&	0.995	\\
3DT-Net ($K=2$)	&	50.26	&	2.13	&	0.50	&	\textbf{0.996}	\\
3DT-Net ($K=3$)	&	\textbf{50.93}	&	\textbf{2.01}	&	\textbf{0.45}	&	\textbf{0.996}	\\
3DT-Net ($K=4$)	&	\underline{50.91}	&	\underline{2.05}	&	\underline{0.46}	&	\textbf{0.996}		\\
  \hline
  \hline
\end{tabular}
% \vspace{-0.2cm}
\normalsize
\end{center}
\end{table}

\subsection{Training Datasets}
We evaluate our proposed method on two publicly simulated hyperspectral imaging datasets including the \emph{CAVE} \cite{yasuma2010generalized}$\footnote{http://www.cs.columbia.edu/CAVE/databases/}$ and the \emph{Harvard} \cite{arad2020ntire}$\footnote{http://vision.seas.harvard.edu/hyperspec/d2x5g3/}$. For real data experiments, we use \emph{WV-2}\footnote{https://www.harrisgeospatial.com/Data-Imagery/Satellite-Imagery/High-Resolution/WorldView-2}.

The CAVE database consists of 32 indoor images captured under controlled illumination with spatial size of $512 \times 512$, including 31 spectral bands of 10 nm wide, covering the visible spectrum from 400 to 700nm. We follow the same setting as \cite{dong2021model}, the HR-MSI (RGB image) is generated by integrating all the ground truth HR-HSI bands with the same simulated spectral response function $R$ of a Nikon D700 camera$\footnote{http://www.maxmax.com/spectral\_response.htm}$, and the LR-HSI can obtained by first applying an anti-aliasing 8$\times$8 Gaussian filter with a standard deviation of 2 to the original HR-HSIs followed by a downsampling along both the horizontal and vertical dimensions with the scaling factor of 8. We select the first 20 samples and randomly extract $64 \times 64$ overlapped patches from them as reference HR-HSIs for training. Thus, the HR-HSIs, HR-MSIs and LR-HSIs are of sizes $64 \times 64 \times 31$, $64 \times 64 \times 3$, $8 \times 8 \times 31$, respectively. The remaining 12 samples of the database are used for testing.

The Harvard database contains 50 indoor and outdoor images recorded under daylight illumination with spatial size of $1,040 \times 1,392$, including 31 spectral bands of 10 nm wide, covering the visible spectrum from 400 to 700nm. Then we use the same method as CAVE database to get the HR-MSI and LR-HSI.  We select the first 30 samples and randomly extract $64 \times 64$ overlapped patches from them as reference HR-HSIs for training. Thus, the HR-HSIs, HR-MSIs and LR-HSIs are of sizes $64 \times 64 \times 31$, $64 \times 64 \times 3$, $8 \times 8 \times 31$, respectively. The remaining 20 samples of the database are used for testing.

The WV-2 database contains a LR-HSI of size $419 \times 658 \times 8$ and an HR-MSI of size $1676 \times 2632 \times 3$, while the ground truth HR-HSI is unavailable. We follow the same setting as \cite{xie2020mhf} to generate the training data. We select the top half part of the HR-MSI $836 \times 2632 \times 3$ and LR-HSI $209 \times 658 \times 8$ as training data and exploit the remaining parts of the data set as testing data. We use the Wald's protocol \cite{zeng2010fusion} to generate the training samples. We divide the training data into $256 \times 256 \times 3$ overlapped HR-MSI patches and $64 \times 64 \times 8$ overlapped LR-HSI patches. Then we down-sampling HR-MSI and LR-HSI patches into size $64 \times 64 \times 3$ and $16 \times 16 \times 8$ as the input HR-MSI and LR-HSI, respectively. Note that the original $64 \times 64 \times 8$ HSI is taken as ground truth.

%------------------------------------------------------------------------
\subsection{Ablation Study}
In this section, we conduct simulated experiments to show the effectiveness of our network in three aspects: the influence of stage number $K$, the effectiveness of 3D-CNN and Transformer prior network, and the effectiveness of 3D-CNN over 2D-CNN.

\setlength{\tabcolsep}{2.750pt}
\begin{table}
%\vspace{1mm}
\begin{center}
%\vspace{1.5mm}
\normalsize
\setlength{\tabcolsep}{4.5pt}
\caption{Ablation study on the prior network with different setting.}\label{tab:prior}
\begin{tabular}{c|ccccc}
  \hline
  \hline
Network	&	Para(M)    &   PSNR	&	SAM	&	ERGAS	&	SSIM\\
\hline
ResNet	& 6.75	&   48.29	&	2.51	&	0.60	&	0.995	\\
U-net	& 6.70	&   49.71	&	2.28	&	0.50	&	\textbf{0.996}	\\
Transformer	& 3.46	&   \underline{50.76}	&	\underline{2.07}	&	\underline{0.46}	&	\textbf{0.996}	\\
3DT-Net	& 3.46	&   \textbf{50.93}	&	\textbf{2.01}	&	\textbf{0.45}	&	\textbf{0.996}	\\
  \hline
  \hline
\end{tabular}
% \vspace{-0.2cm}
\normalsize
\end{center}
\end{table}

\begin{figure*}[t] %H为当前位置，!htb为忽略美学标准，htbp为浮动图形
\centering %图片居中
\includegraphics[width=0.99\textwidth]{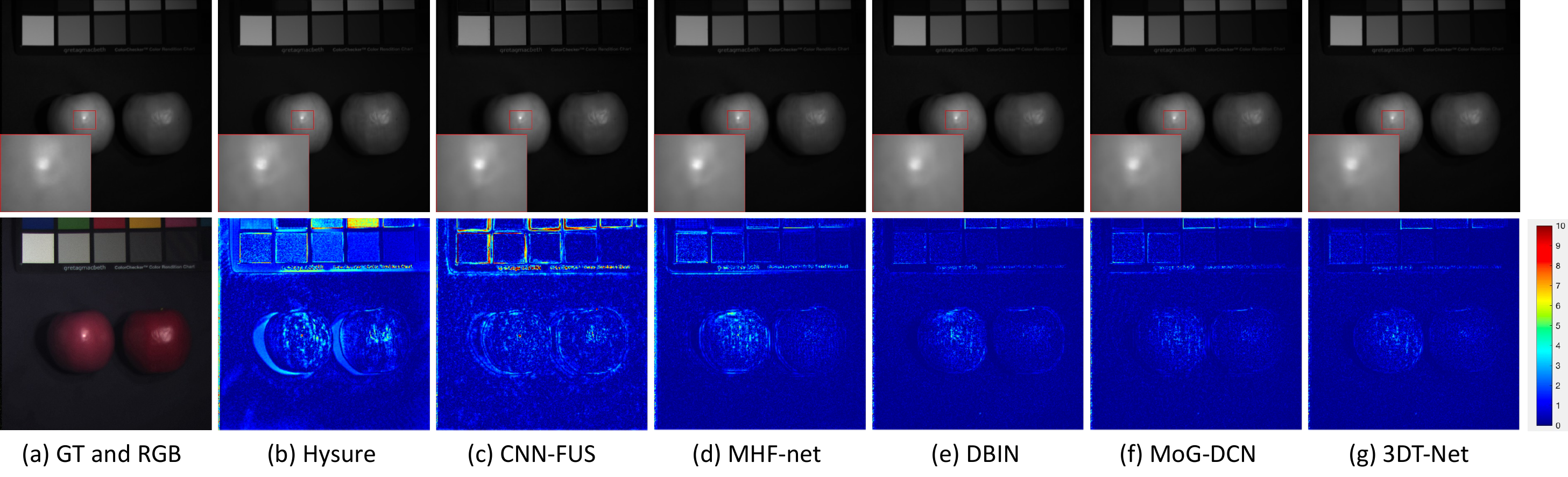} %插入图片，[]中设置图片大小，{}中是图片文件名
\caption{Qualitative results of the CAVE datasets at band 31. Top row: Ground truth and reconstructed images by 6 comparison methods, with a demarcated areas zoomed in 4 times for easy observation. Bottom row: RGB and the error images of the result obtain by the 6 competing methods.} %最终文档中希望显示的图片标题
\label{fig:3DT-Net_cave} %用于文内引用的标签
\vspace{-0.2cm}
\end{figure*}

\noindent\textbf{Effect of stage number $K$.} To explore the impact of the number of unfolded stages on the HSISR performance, we report the performance of the proposed 3DT-Net with different stage number $K$. Table~\ref{tab:iterations} shows the average results over 12 testing HSIs. Here, the bold values represent the best result, and the results with underlines denote the second best. From the table, we can observe that as the number of iterations increases, the performance of PSNR and SSIM will increase, but when $K \textgreater 3$, the performance of PSNR and SSIM will decrease. This shows that our method can quickly converge after several iterations, but the performance may not be improved when the iteration is increased, because the network is too complex and it will be difficult to train to obtain the optimal solution. In the following experiments, we choose $K=3$ in our implementation.

\begin{table}
%\vspace{1mm}
\begin{center}
%\vspace{1.5mm}
\normalsize
\setlength{\tabcolsep}{6.5pt}
\caption{Quantitative comparisons of different approaches over 10 testing images from the CAVE dataset with respect to four PQIs.}\label{tab:CAVE}
\begin{tabular}{c|ccccc}
  \hline
  \hline
Method	&	PSNR	&	SAM	&	ERGAS	&	SSIM\\
\hline
Hysure \cite{simoes2014convex}	&	40.06	&	9.66	&	1.30	&	0.976	\\
NSSR \cite{dong2016hyperspectral}	&	44.07	&	4.35	&	0.82	&	0.987	\\
CSTF \cite{li2018fusing}	&	42.41	&	5.04	&	0.87	&	0.979	\\
LTTR \cite{dian2019learning}	& 	45.89	&	2.97	&	0.66	&	0.994		\\
CNN-FUS \cite{dian2020regularizing}	& 	44.21	&	4.04	&	0.82	&	0.989	\\
MHF-net \cite{xie2020mhf}	&	46.31	&	3.39	&	0.64	&	0.994\\
DBIN \cite{wang2019deep}	& 48.82	&	2.09	&	0.50	&	\textbf{0.996}	\\
MoG-DCN \cite{dong2021model} &	\underline{49.89}	&	\underline{2.04}	&	\textbf{0.45}	&	\textbf{0.996}	\\
3DT-Net	& 	\textbf{50.93} &	\textbf{2.01}	&	\textbf{0.45}	&	\textbf{0.996}\\
  \hline
  \hline
\end{tabular}
\vspace{-0.2cm}
\normalsize
\end{center}
\end{table}

\noindent\textbf{Effect of 3D-CNN and Transformer prior network.} To verify the effectiveness of 3D-CNN and Transformer prior network, we use the ResNet prior network adopted in \cite{xie2020mhf, wang2019deep} and U-net prior network adopted in \cite{dong2021model} to replace the proposed 3D-CNN and Transformer prior network, respectively. Considering the powerful long-range modeling ability of Transformer, it only needs very few parameters to achieve strong performance. We use the U-net prior network proposed in \cite{dong2021model} as the best U-net prior network (its parameter amount may lager than our method), rather than requiring that the parameters of the U-net prior network must be the same as these of the 3D-CNN and Transformer prior network. For fair comparison of ResNet, we let the parameters of ResNet prior network the same as the U-net prior network. From Table~\ref{tab:prior}, we can see that the PSNR value of Transformer prior network is 1.05 dB higher than that of the U-net prior network, and the parameters are only half of the U-net prior network. After the introduction of 3D convolutional layers (the only difference between Transformer and 3DT-Net), the PSNR performance increased again by 0.17 dB, which proves that 3D convolutional layers can better extract the spatio-spectral correlation priors.

\begin{figure*}[t] %H为当前位置，!htb为忽略美学标准，htbp为浮动图形
\centering %图片居中
\includegraphics[width=0.99\textwidth]{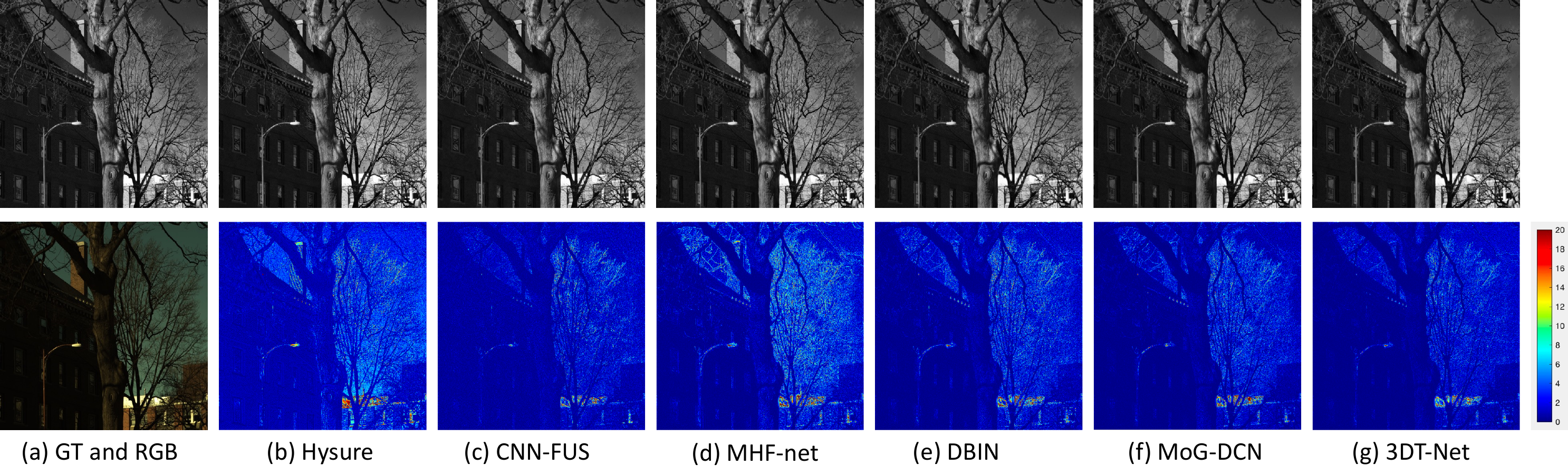} %插入图片，[]中设置图片大小，{}中是图片文件名
\caption{Qualitative results of the Harvard datasets at band 31. Top row: Ground truth and reconstructed images by 6 comparison methods. Bottom row: RGB and the error images of the result obtain by the 6 competing methods.} %最终文档中希望显示的图片标题
\label{fig:3DT-Net_harvard} %用于文内引用的标签
% \vspace{-0.2cm}
\end{figure*}

\begin{table}
%\vspace{1mm}
\begin{center}
%\vspace{1.5mm}
\normalsize
\setlength{\tabcolsep}{6.5pt}
\caption{Quantitative comparisons of different approaches over 20 testing images from the Harvard dataset with respect to four PQIs.}\label{tab:Harvard}
\begin{tabular}{c|ccccc}
  \hline
  \hline
Method	&	PSNR	&	SAM	&	ERGAS	&	SSIM\\
\hline
Hysure \cite{simoes2014convex}	&   44.26	&	3.75	&	1.40	&	0.983	\\
NSSR \cite{dong2016hyperspectral}	&  46.08	&	3.40	&	1.20	&	0.985	\\
CSTF \cite{li2018fusing}	&   44.98	&	3.54	&	1.07	&	0.980	\\
LTTR \cite{dian2019learning}	&  46.86	&	2.90	&	1.11	&	0.987		\\
CNN-FUS \cite{dian2020regularizing}	&   46.05	&	3.24	&	1.12	&	0.985	\\
MHF-net \cite{xie2020mhf}	&   46.42	&	3.01	&	1.09	&	0.987\\
DBIN \cite{wang2019deep}	&  47.36	&	2.71	&	0.97	&	\underline{0.988}	\\
MoG-DCN \cite{dong2021model} &  	\underline{47.64}	&	\underline{2.67}	&	\underline{0.91}	&	\underline{0.988}	\\
3DT-Net	&   	\textbf{48.05}	&	\textbf{2.27}	&	\textbf{0.84}	&	\textbf{0.991}\\
  \hline
  \hline
\end{tabular}
\vspace{-0.2cm}
\normalsize
\end{center}
\end{table}

\begin{figure*}[t] %H为当前位置，!htb为忽略美学标准，htbp为浮动图形
\centering %图片居中
\includegraphics[width=0.99\textwidth]{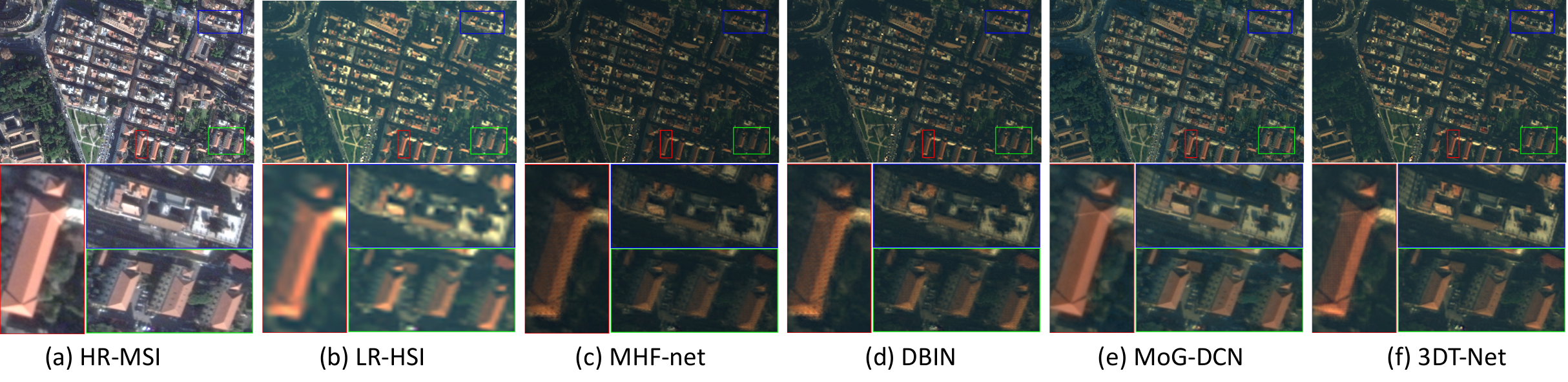} %插入图片，[]中设置图片大小，{}中是图片文件名
\caption{Qualitative results of the WV-2 dataset. (a) and (b) are the HR-MSI and LR-HSI of the right bottom area of Roman Colosseum. We show the composite image of the HSI with bands 5-3-2 as R-G-B. (c)-(f) reconstructed images by 6 comparison methods, with a demarcated areas zoomed for easy observation.} %最终文档中希望显示的图片标题
\label{fig:WV2result} %用于文内引用的标签
% \vspace{-0.2cm}
\end{figure*}

%------------------------------------------------------------------------
\subsection{Experiments with Simulated Data}
The average results in terms of the PSNR, SAM, ERGAS and SSIM of all the competing methods on the CAVE and Harvard datasets are reported in Table~\ref{tab:CAVE} and Table~\ref{tab:Harvard}, respectively. From these two tables, it can be seen that the performance of deep learning-based methods are better than that of traditional shallow learning-based methods, which shows the advantages and representation ability of deep neural networks. Clearly, the proposed 3DT-Net method outperforms all other competing methods by a considerable margin. The average PSNR value of our method is more than 1.0 dB and 0.4 dB higher than that of the second best method (\emph{i.e.}, the MoG-DCN method) on CAVE and Harvard datasets, respectively.

In Fig. \ref{fig:3DT-Net_cave} and Fig. \ref{fig:3DT-Net_harvard}, we show parts of the reconstructed HR-HSIs at 700 nm by the competing methods for the test image \textsl{real and fake apples} from the CAVE dataset and the test image \textsl{imgf2} from the Harvard dataset, respectively. From Fig. \ref{fig:3DT-Net_cave}, we can see that the composite image obtained by 3DT-Net is the closest to the ground-truth, while the results of Hysure and CNN-FUS contain obvious incorrect structure or spectral distortion. From Fig. \ref{fig:3DT-Net_harvard}, we can see that all the test methods can well reconstruct the HR spatial structures of the HSI. Obviously, the proposed method performs best in recovering the details of the original HSI and achieves the smallest reconstruction error.

\subsection{Experiments with Real Data}
To verify the robustness of our method on real data, a public dataset of real MSIs called WV-2 are used in our experiments. In Fig. \ref{fig:WV2result}, we show a portion of the fusion result of the testing data. It can be seen that MHF-net \cite{xie2020mhf} and DBIN \cite{wang2019deep} have obvious blur and artifacts. Visual inspection clearly shows that the proposed method has better visual effect.

%------------------------------------------------------------------------
\section{Conclusions}
\label{sec:Conclusions}
In this paper, we have provided a new deep unfolding network for hyperspectral image super-resolution based on proximal gradient algorithm and Transformer prior. Unlike other deep learning-based HSI super-resolution methods using CNN to learn the prior of HSIs, our 3DT-Net use Transformer layers to exploit spatial priors and 3D convolutional layers to exploit spatio-spectral correlation priors. Compared with CNN, Transformer has long-range modeling capabilities and can use fewer parameters to achieve better performance. 3D convolution slides along the spatial and channel dimensions at the same time, making it more suitable for exploring the spatio-spectral correlation of HSIs. Evaluations on two public simulated datasets and one real datasets both demonstrate that the proposed 3DT-Net achieves state-of-the-art performance in terms of quantitative result and visual quality.

The proposed prior modeling method is general for many low-level image restoration tasks. In particular, it can be seamlessly integrated with any method under the Plug-and-Play framework, especially suitable for processing high-dimensional data, such as hyperspectral image, video, and light-field. The limitation of our approach is that it is a little more computationally taxing than those convolutional residual block based network. Therefore, in the future we will exploit some efficient implementations to extend our method to other tasks.

%%%%%%%%% REFERENCES
{\small
\bibliographystyle{ieee_fullname}
\bibliography{egbib}
}

\end{document}